\documentclass[%
 aip,
 amsmath,amssymb,
 reprint,%
]{revtex4-1}

\usepackage{graphicx}
\usepackage{dcolumn}
\usepackage{bm}

\usepackage[utf8]{inputenc}
\usepackage[T1]{fontenc}
\usepackage{mathptmx}
\usepackage{etoolbox}
\usepackage{amsmath}
\usepackage{algorithm}
\usepackage{algpseudocode}
\usepackage{xcolor}

\makeatletter
\def\@email#1#2{%
 \endgroup
 \patchcmd{\titleblock@produce}
  {\frontmatter@RRAPformat}
  {\frontmatter@RRAPformat{\produce@RRAP{*#1\href{mailto:#2}{#2}}}\frontmatter@RRAPformat}
  {}{}
}%
\makeatother
\begin{document}

\preprint{AIP/123-QED}

\title{Considerations for extracting moir\'{e}-level strain from dark field intensities in transmission electron microscopy} 

\author{Isaac M. Craig}
\affiliation{Department of Chemistry, University of California, Berkeley, California, 94720, USA}
\author{Madeline Van Winkle}
\affiliation{Department of Chemistry, University of California, Berkeley, California, 94720, USA}
\author{Colin Ophus}
\affiliation{The Molecular Foundry, Lawrence Berkeley National Laboratory, Berkeley, California, 94720, USA}
\author{D. Kwabena Bediako}
\email[]{bediako@berkeley.edu}
\affiliation{Department of Chemistry, University of California, Berkeley, California, 94720, USA}
\affiliation{Chemical Sciences Division, Lawrence Berkeley National Laboratory, California, 94720, USA}

\date{\today}

\begin{abstract}
Bragg interferometry (BI) is an imaging technique based on four-dimensional scanning electron microscopy (4D-STEM) wherein the intensities of select overlapping Bragg disks are fit or more qualitatively analyzed in the context of simple trigonometric equations to determine local stacking order. In 4D-STEM based approaches, the collection of full diffraction patterns at each real-space position of the scanning probe allows the use of precise virtual apertures much smaller and more variable in shape than those used in conventional dark field imaging, such that even buried interfaces marginally twisted from other layers can be targeted. A coarse-grained form of dark field ptychography, BI uses simple physically derived fitting functions to extract the average structure within the illumination region and is therefore viable over large fields of view. BI has shown a particular advantage for selectively investigating the interlayer stacking and associated moir\'{e} reconstruction of bilayer interfaces within complex multi-layered structures. This has enabled investigation of reconstruction and substrate effects in bilayers through encapsulating hexagonal boron nitride and of select bilayer interfaces within trilayer stacks. However, the technique can be improved to provide a greater spatial resolution and probe a wider range of twisted structures, for which current limitations on acquisition parameters can lead to large illumination regions and the computationally involved post-processing can fail. Here we analyze these limitations and the computational processing in greater depth, presenting a few methods for improvement over previous works, discussing potential areas for further expansion, and illustrating the current capabilities of this approach for extracting moir\'{e}-scale strain. 
\end{abstract}

\maketitle

\section{Introduction}
Moir\'{e} materials, formed by stacking atomically thin materials with a slight interlayer twist or lattice mismatch, exhibit spatially modulated potentials that can localize excitons and charge carriers, serving as a basis for many novel devices. \cite{huang2022excitons,mak2022semiconductor,zhang2018moire,seyler2019signatures,alexeev2019resonantly,jin2019observation,tran2019evidence,liu2021signatures,dandu2022electrically,naik2022intralayer,susarla2022hyperspectral} Both highly tunable and proposed to be a physical realization of the Hubbard \color{black} and other correlated models \color{black} with experimentally accessible phase transitions, moir\'{e} materials are also a preferred platform to investigate \color{black} and understand correlated electron physics outside of computationally tractable regimes. \cite{balents2020superconductivity,tang2020simulation,wang2020correlated,regan2020mott,xu2020correlated,li2021imaging,huang2021correlated,xu2022tunable, li2021quantum,dean2013hofstadter,kang2024evidence,serlin2020intrinsic,chang2023colloquium,xie2022valley,zhang2020quantum,xu2023observation,shi2021moire,kennes2021moire,chen2020tunable,lu2024fractional,pan2020quantum,park2023observation,wang2015evidence} \color{black} However as these materials locally reconstruct \cite{yoo2019atomic,weston2020atomic,rosenberger2020twist,kazmierczak2021strain,li2021imaging2, sung2022torsional} in a manner that alters their electronic behavior \cite{shabani2021deep, naik2020origin, li2021lattice, naik2018ultraflatbands,enaldiev2020stacking, ferreira2021band}, a detailed understanding of the reconstruction mechanism in these systems is crucial to realize practical devices and further our understanding of their rich physics.

We discuss an interferometric methodology based on four-dimensional scanning transmission electron microscopy \cite{ophus2014recording,ophus2019four} that permits the direct measurement of inter-layer strain fields. \cite{kazmierczak2021strain, van2023rotational} This approach involves fitting the intensity interference \cite{Latychevskaia2018Holography} of overlapping Bragg disks in the diffraction patterns to a known functional form to extract structural details. \cite{kazmierczak2021strain, zachman2021interferometric, van2023rotational} Alternative 4D-STEM based methods typically fit the locations of Bragg disks \cite{mukherjee2020lattice, yuan2019lattice, munshi2022disentangling, sari2023analysis, mahr2021accurate}, or have analyzed a combination of both intensity and disk locations \cite{shi2023domain}. However, determining the locations of these large converged beam electron diffraction (CBED) disks to the precision needed for revolving moir\'{e}-level strain is difficult without specialized apertures \cite{zeltmann2020patterned} and has not yet been feasible in marginally twisted structures. \color{black} Previous works have also analyzed the Bragg disk overlap intensities in single-shot CBED patterns, \cite{latychevskaia2018convergent, latychevskaia2020convergent, latychevskaia2021holographic, latychevskaia2020holographic} both for the focused probes we consider and with large defocus values where holographic fringes encoding out-of-plane information are seen. However, these works had not yet fit these intensities to extract local structure within a scanning set-up, as is more characteristic of ptychography. \color{black} 

Other ptychographic techniques \color{black} used in electron microscopy similarly analyze changes in the scattering intensities when scanning a converged probe \color{black} and can provide much higher resolution \cite{jiang2018electron, chen2020mixed, yang2024local, pelz2017low}, but are currently arduous to acquire and analyze over the moir\'{e} length scale. BI avoids the same level of computational complexity by only fitting select areas in the acquired diffraction patterns of anticipated interest for determining moir\'{e} stacking to simple physically derived analytical expressions, omitting the central beam and averaging over the sample within the illuminated region to provide a coarse-grained picture of reconstruction on the moir\'{e}-scale. \color{black} We refer to this heavily constrained analog of dark field ptychography, realized in x-ray diffraction as Bragg ptychography\cite{hruszkewycz2017high,pfeiffer2018x,takahashi2013bragg,kim2018three}, as Bragg interferometry \cite{kazmierczak2021strain} to reflect the lack of an attempt to reconstruct the probe wavefunction.\color{black}

We outline how this interferometric methodology can measure strain from bilayer interfaces in generic structures and discuss and present solutions to various numerical challenges encountered in its practical application. Numerical approaches for the fitting and extraction of strain are compared, and various considerations are presented for improved strain resolution and analyzed using simulated multi-slice data and through re-processing published \cite{kazmierczak2021strain, van2023rotational} data. 

\section{Strain Fields in Moir\'{e} Systems}

We begin with a simple picture of strain in moir\'{e} materials. We can first assume that there is negligible hetero-strain within the moir\'{e} such that the lattice vectors of both layers are identical apart from a small interlayer twist angle $\theta_t$ (Figure 1a), resulting in the following relationship between the atomic positions of the top layer $\textbf{r}_{xy}^{\rm{top}}$, the bottom layer $\textbf{r}_{xy}^{\rm{bottom}}$, and the reference layer $\textbf{r}_{xy}^{\rm{ref}}$ we take as halfway between them at each pixel location $(x,y)$. The details of the $\textbf{r}_{xy}$ coordinates do not factor into this analysis as we consider the strain of the locally averaged offset between the lattice vectors of each layer.

\begin{align*}
\textbf{r}_{xy}^{\rm{top}} = 
\begin{bmatrix}
 \cos(\frac{\theta_t}{2})  & -\sin(\frac{\theta_t}{2}) \\
 \sin(\frac{\theta_t}{2}) &\cos(\frac{\theta_t}{2}) \\
\end{bmatrix}
\textbf{r}_{xy}^{\rm{ref}}
\end{align*}

\begin{align*}
\textbf{r}_{xy}^{\rm{bottom}} = 
\begin{bmatrix}
 \cos(\frac{\theta_t}{2})  & \sin(\frac{\theta_t}{2}) \\
 -\sin(\frac{\theta_t}{2}) &\cos(\frac{\theta_t}{2}) \\
\end{bmatrix}
\textbf{r}_{xy}^{\rm{ref}}
\end{align*}

The expression is straightforwardly adapted to account for an interlayer lattice mismatch originating either in hetero-strain or from the use of two chemically distinct layers. The intralayer displacement $\textbf{u}_{xy}^{\rm{top}}$ (Fig 1b) associated with the displacement of each atom in the top layer away from the reference is therefore given by $\textbf{u}^{\rm{top}}_{xy} = (\textbf{r}_{xy}^{\rm{top}} - \textbf{r}_{xy}^{\rm{ref}})$, yielding the following relation. 

\begin{align*}
\textbf{u}_{xy}^{\rm{top}} = 
\begin{bmatrix}
 \cos(\frac{\theta_t}{2}) -1  & -\sin(\frac{\theta_t}{2}) \\
 \sin(\frac{\theta_t}{2}) &\cos(\frac{\theta_t}{2}) -1 \\
\end{bmatrix}
\textbf{r}_{xy}^{\rm{ref}}
\end{align*}

\begin{figure}
\includegraphics[scale=0.80]{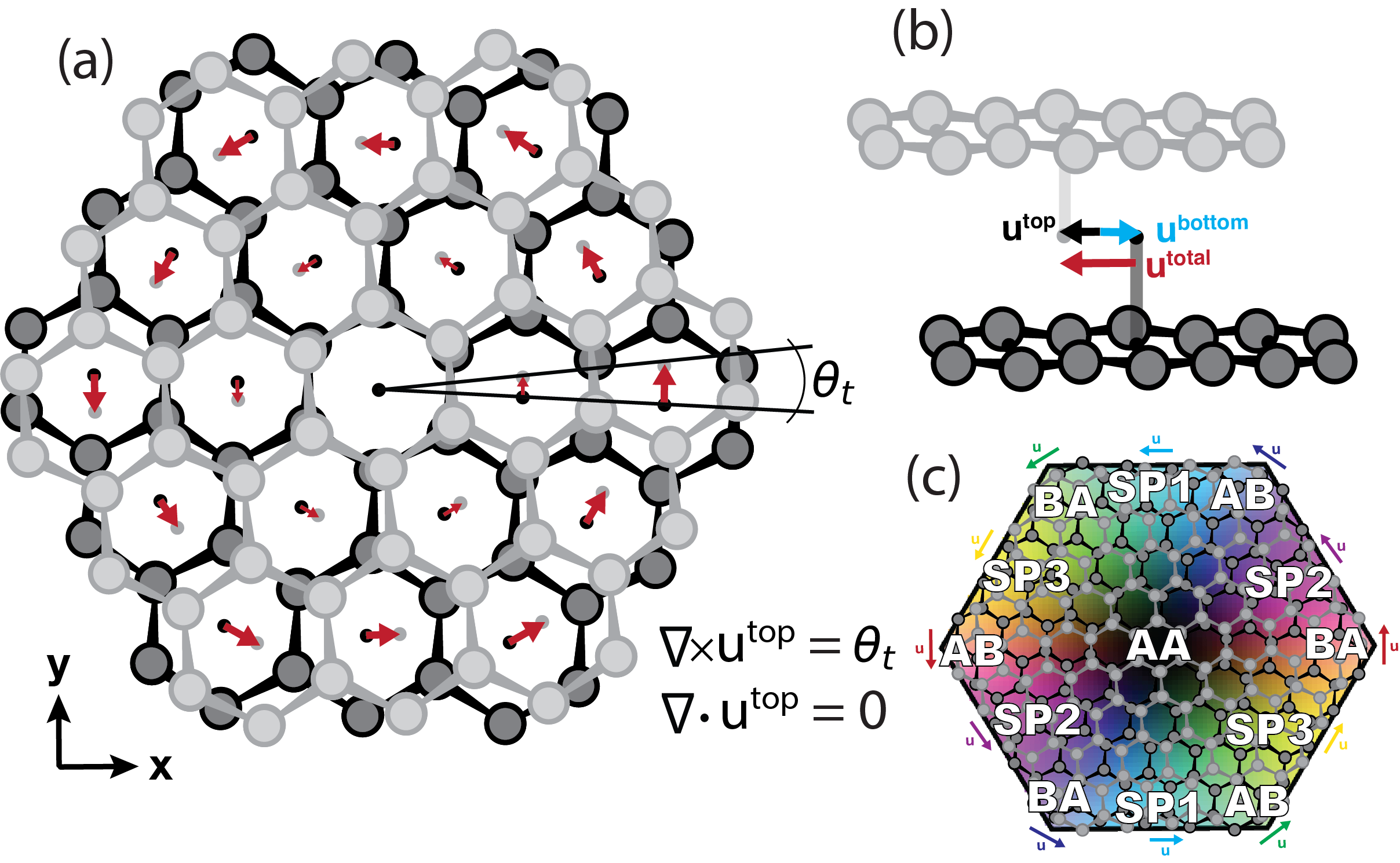}
\caption{ \textbf{(1a)} Schematic illustrating the anticipated interlayer displacements $\textbf{u}_{xy}^{\rm{total}}$ within a rigid moir\'{e} structure at a twist angle of $\theta_t$, for which the curl and divergence of the displacement $\textbf{u}_{xy}^{\rm{top}}$ (equal to one half $\textbf{u}_{xy}^{\rm{total}}$) are uniformly $\theta_t$ and 0 respectively. \textbf{(1b)} Definition of the total interlayer displacement $\textbf{u}_{xy}^{\rm{total}}$, where $\textbf{u}_{xy}^{\rm{total}}$ = $\textbf{u}_{xy}^{\rm{top}}$ and $\textbf{u}_{xy}^{\rm{top}}$ - $\textbf{u}_{xy}^{\rm{bottom}}$ in terms of the intralayer displacements of each layer, $\textbf{u}_{xy}^{\rm{top}}$ and $\textbf{u}_{xy}^{\rm{bottom}}$. |$\textbf{u}_{xy}^{\rm{top}}$| $\approx$ |$\textbf{u}_{xy}^{\rm{bottom}}$| such that $\textbf{u}_{xy}^{\rm{total}}$ $\approx 2\textbf{u}_{xy}^{\rm{top}}$ is assumed. \textbf{(1c)} Legend illustrating the coloring scheme used to denote the atomic stacking, where AA denotes regions where the two layers are aligned out-of-plane and AB/BA denote regions where the two layers are aligned Bernal out-of-plane. SP1, SP2, and SP3 denote the various symmetrically distinct saddle point stacking regions. }
\end{figure}

Likewise, we define  $\textbf{u}_{xy}^{\rm{bottom}} = (\textbf{r}_{xy}^{\rm{bottom}} - \textbf{r}_{xy}^{\rm{ref}})$. These displacements can be analyzed using small displacement theory (also known as infinitesimal strain theory) \cite{slaughter2002chapter}, in which the strain tensors are linearized. This framework neglects higher-order terms from changes to the coordinate system, omitting the distinction between deformed and reference axes. Such an approximation is valid when the displacements change relatively slowly over the field of view such that $|\nabla \textbf{u}_{xy}^{\rm{top}}|_{\infty} \ll 1$. This necessitates that the relaxation is not too dramatic and that the lattice scale $a_0$ is much smaller than the moir\'{e}-wavelength, coinciding with a small interlayer twist. More precisely, the local deformation gradient within a rigid structure is on the order of $\sin(\theta/2)$ as $\textbf{u}^{\rm{top}}_{xy}$ varies between 0 and $a_0/4$ over a length scale $a_0/\sin(\theta/2)$. The following analysis therefore assumes $\sin(\theta/2) \ll 1$.

\begin{figure*}
\includegraphics[scale=0.85]{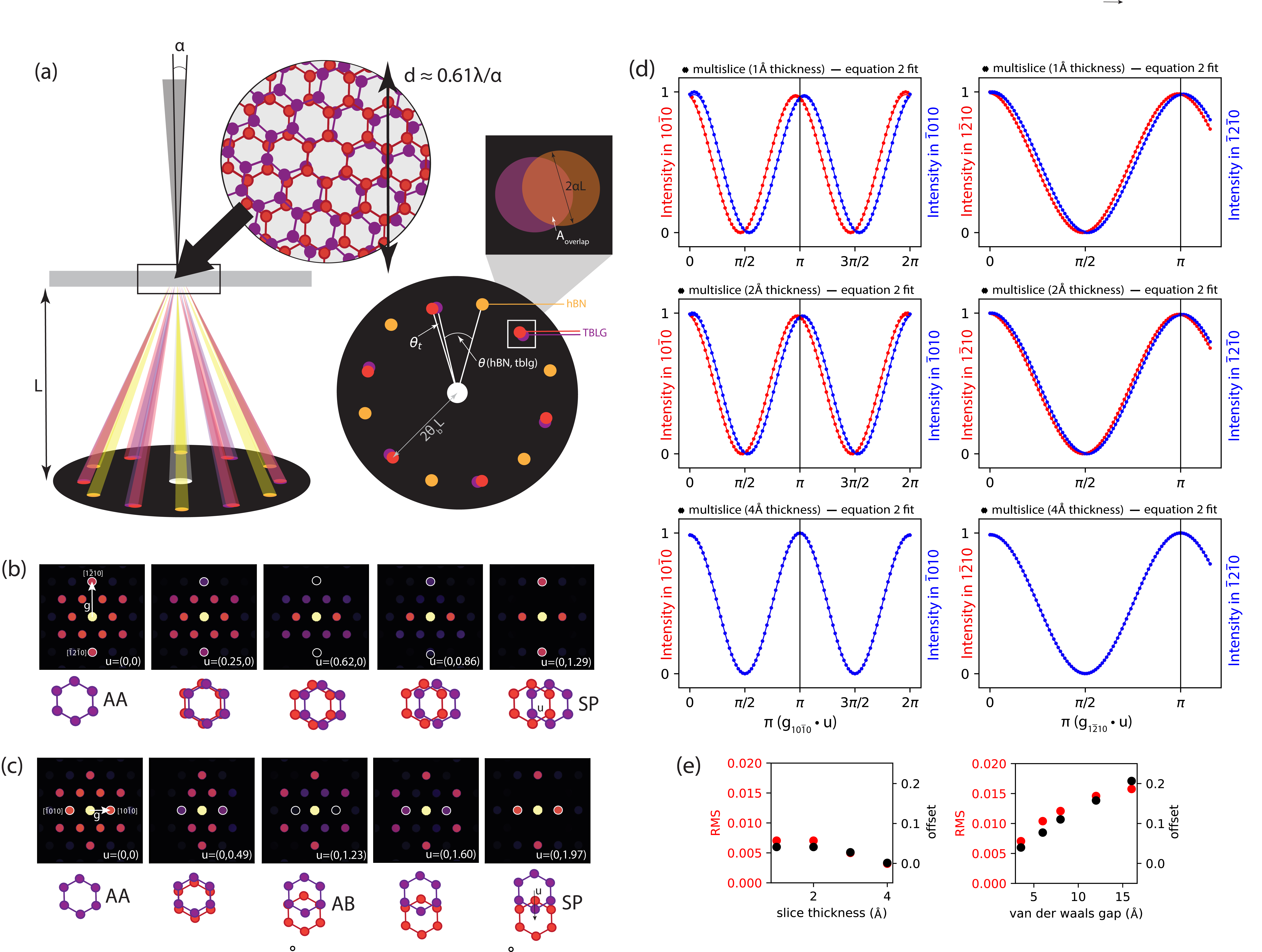}
\caption{ \textbf{(2a)} Schematics illustrating the 4D-STEM Bragg interferometry approach where a series of converged beam electron diffraction patterns are collected at various real-space positions as the beam is scanned across the sample. Various relevant parameters including the camera length L, convergence semi-angle $\alpha$, and electron probe width d are noted. \textbf{(2b-c)} Multi-slice CBED patterns at select $\textbf{u}_{xy}$ choices. Corresponding approximate real space stackings are below. $10\bar{1}0$ and $\bar{2}110$ reflections are highlighted in white, corresponding to those plotted in 2d. \textbf{(2d)} Variation in intensity of the $10\bar{1}0$ and $1\bar{2}10$ reflections (red) and their Friedel pairs (blue) over a densely-sampled scan of $\textbf{u}_{xy}$ choices. These reflections were selected for ease as they align with the chosen cartesian axis orientation. Both intensities show good agreement with a fit to equation 2 (solid lines) with the need for a small offset from the origin, equal and opposite for each Friedel pair. \textbf{(2e)} Corresponding root mean squared errors and offsets added to the cosine argument are shown for the $10\bar{1}0$ line-cut (left column of 2d) in units of $\pi$ for a series of acquisition parameters. Unless otherwise noted, multi-slice simulations were carried out using $\alpha$ = 4 mrad, a slice thickness of 3 \AA, perfect aberration correction, and the VdW gap expected for graphene of $3.55$\textrm{\AA} using AbTEM \cite{madsen2021abtem} }  
\end{figure*}

The strain tensor $\nabla\textbf{u}_{xy}^{\rm{top}}$ then consists of partial derivatives of $\textbf{u}_{xy}^{\rm{top}}$ with respect to the reference coordinates. For a rigidly twisted structure that perfectly obeys this relation at each pixel, $\nabla\textbf{u}_{xy}^{\rm{top}}$ is independent of pixel location and obeys the following uniformly. 

\begin{align}
\nabla \textbf{u}_{xy}^{\rm{top}} 
=  
\begin{bmatrix}
 \cos(\frac{\theta_t}{2}) -1  & -\sin(\frac{\theta_t}{2}) \\
 \sin(\frac{\theta_t}{2}) &\cos(\frac{\theta_t}{2}) -1 \\
\end{bmatrix}
\end{align}

This rigid displacement field corresponds to a uniform local curl and divergence of $\nabla \times \textbf{u}_{xy}^{\rm{top}} = 2\sin (\theta_t/2)$ and  $\nabla \cdot \textbf{u}_{xy}^{\rm{top}} = 2\cos(\theta_t/2) - 2$, equal to $\theta_t$ and 0 respectively to first order. Atomic reconstruction will manifest deviations from this uniform curl and divergence, which we will denote as rotations and dilations respectively. The use of small deformation theory further permits us to consider reconstruction-driven changes to the local rotation and dilation as additive changes to what is expected of a rigid moir\'{e}. In this framework, we can decompose $\nabla \textbf{u}_{xy}^{\rm{top}}$ as the sum of the curl-free and divergence-free matrices $\epsilon_{xy}$ and $\omega_{xy}$, encoding dilational and rotational strain respectively.

\begin{align*}
\nabla \textbf{u}_{xy}^{\rm{top}}
=  
\underbrace{ \frac{1}{2} \nabla \textbf{u}_{xy}^{\rm{top}} + \frac{1}{2} (\nabla \textbf{u}_{xy}^{\rm{top}})^T }_{\epsilon_{xy}} + 
\underbrace{ \frac{1}{2} \nabla \textbf{u}_{xy}^{\rm{top}} - \frac{1}{2} (\nabla \textbf{u}_{xy}^{\rm{top}})^T }_{\omega_{xy}}
\end{align*}

The eigenvalues and eigenvectors of $\epsilon$ are termed the principal stretches and their directions, and the sum and difference of these eigenvalues are the total intralayer dilation (the first invariant, equivalent to $\nabla \cdot \textbf{u}_{xy}^{\rm{top}}$) and maximum intralayer engineering shear strain $\gamma_{xy}$ respectively. The curl of $\omega$ corresponds to the local rotation $\nabla \times \textbf{u}_{xy}^{\rm{top}}$, consisting of the sum of the rigidly imposed twist angle and a local reconstruction rotation. For a homo-bilayer structure, the deformation of the top and bottom layers are equal and opposite and all reconstruction information can be determined through consideration of a single layer. In the discussion that follows, we use $\textbf{u}_{xy}$ to denote $\textbf{u}_{xy}^{\rm{total}}$.

\section{Interference Region Intensities for Displacement Measurements}

To extract these displacements $\textbf{u}_{xy}$ and associated strain $\nabla \textbf{u}_{xy}$, we look at the intensity of the overlap regions between Bragg disks arising from the scattering off of desired layers in the material. The overlap intensities $I_i$ at reciprocal lattice positions $\textbf{g}_i$ can be shown to obey the following equation, where $\textbf{u}_{xy}$ is the (in-plane projection of the) interlayer displacement vector dictating stacking order. 

\begin{eqnarray}
I_i &= A_i + B_i cos^2(\pi \textbf{g}_i \cdot \textbf{u}_{xy}) 
\end{eqnarray}

The full derivation is shown in the supplemental information of \cite{kazmierczak2021strain} and \cite{van2023rotational} for a centro-symmetric homo-bilayer and a general bilayer respectively. Briefly, a series of strict assumptions (that the beam is completely focused and that scattering takes place within a single plane and is well represented by the weak phase object approximation) are used to yield the simple analytic expression above. \color{black} We also note that an equivalent expression for centrosymmetric materials may be obtained by taking the $\Delta f = 0$ limit of functional forms derived for in-line CBED holography. \cite{latychevskaia2018convergent, latychevskaia2020convergent, latychevskaia2021holographic, latychevskaia2020holographic} One possible avenue for increased generalization of these fitting functions may use the anticipated z-dependence of these expressions for $\Delta f \neq 0$ to additionally measure corrugations, however we note that large defocus values strongly impact the subsequent discussion of resolution so such an approach is difficult in practice. \cite{zachman2021interferometric}\color{black}  


The accuracy of this expression can be investigated through comparison to simulated electron diffraction patterns, obtained using the multi-slice algorithm. \cite{cowley1957scattering} In figure 2, we present multi-slice results obtained at a series of van-der waals (VdW) gap choices and multi-slice thicknesses, parameterizing the number of scattering events. Each multi-slice simulation is performed on a large uniform bilayer graphene sheet of a given stacking order to investigate the accuracy of equation (2) without resolution considerations, which will be discussed in a later section. An $\alpha$ of 4 mrad and a perfectly focused aberration-free lens are assumed. We note that the primary effect of changing $\alpha$ in the absence of resolution considerations will be to increase the impact of small aberrations not included here. Aberrations will also change the internal structure of the overlap intensity, effects from which can be mitigated through averaging over the whole overlap region. The implications of subtle internal structure in abberated probes for non-spherical virtual apertures is outside the scope of this paper. The defocus also will have little impact when resolution considerations are ignored so long as the entire overlap region intensity is averaged \cite{zachman2021interferometric}. 

Fig. 2d shows a slight disagreement between Eqn 2. and the multi-slice results. This discrepancy presents as a slight decrease in the multi-slice intensity of the second-order reflections in the AA-type stacking region and a small offset of the cosine origin. Increasing the VdW gap of the material leads to a greater deviation from the expected expression, as it is derived under the assumption that the free-space propagation present in multi-slice between the two layers plays a minor role. However, the expression remains accurate up to at least gaps of 15 angstroms. We note that these discrepancies are far too small to be seen in the experimental data due to the effects of noise and finite resolution for materials with relatively small VdW gaps, but may need to be considered in bilayer materials with many intermediate layers between.

The slight discrepancy between equation (2) and the multislice results obtained using a slice thickness greater than the VdW gap (and therefore a single scattering event) can be attributed to the weak phase object assumption. The goodness of fit for the data simulated with a 4\rm{\AA} slice thickness reflects that the weak phase object assumption is well motivated for bilayer graphene and slight deviations from equation (2) are primarily driven by multiple scattering. We conclude that the presented expression is sufficient for describing the intensity variations in graphene bilayers and inaccuracies arising from finite spatial resolution, post-processing, and experimental considerations such as sample purity will dominate.

\begin{figure*}
\includegraphics[scale=0.9]{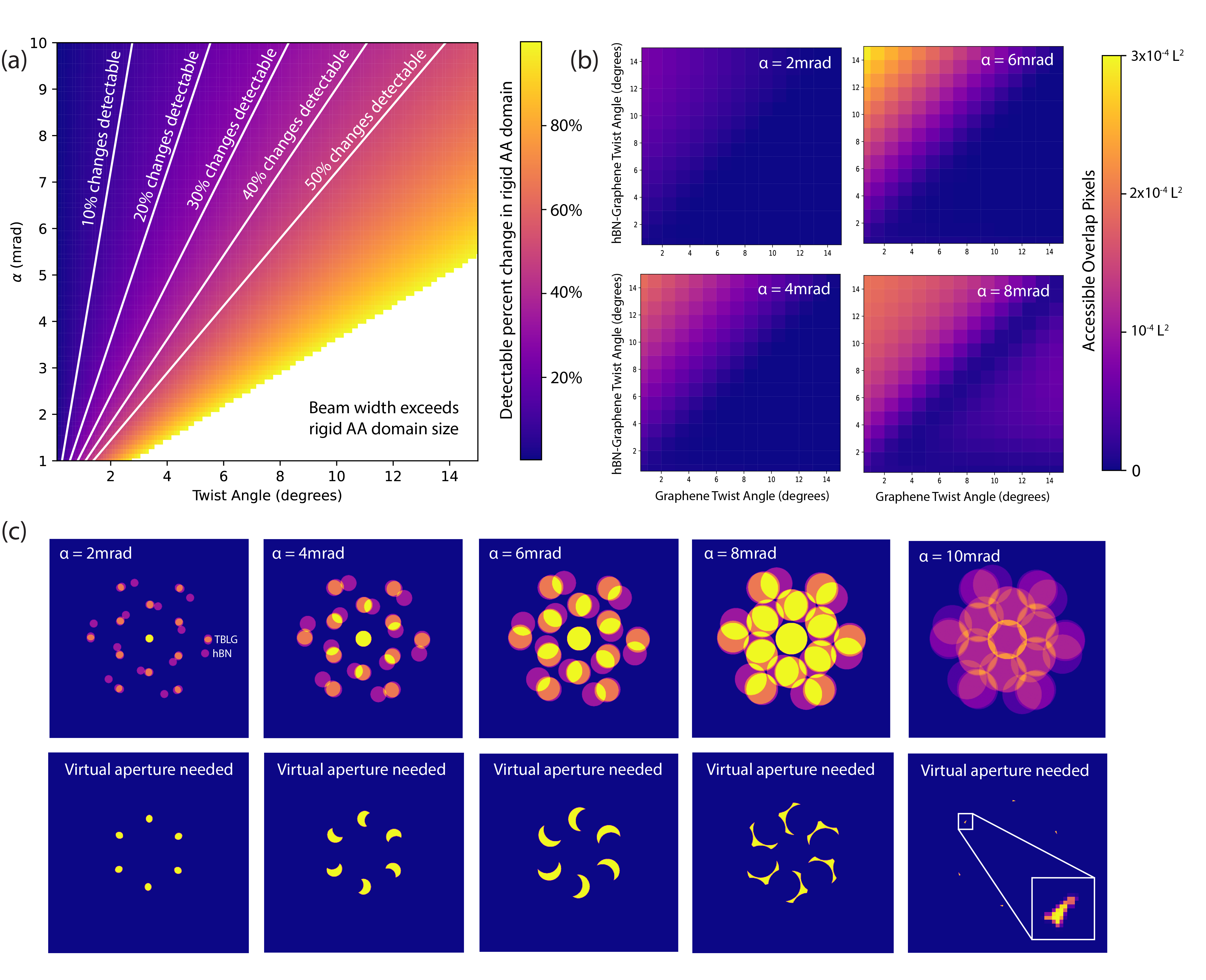}
\caption{ \textbf{(3a)} Convergence semi-angle $\alpha$ needed to resolve AA-type domain contractions to a given precision. The probe size is assumed to be 0.61 $\lambda/\alpha$, AA region defined as $|u| = 0 \pm 0.25 a_0$, and domain size changes are taken to be observable when larger than the beam width. The $\alpha$ needed will be over-estimated as observed beam widths are slightly smaller than those predicted and many independent measurements are used. \textbf{(3b)} Number of viable overlap pixels in units of the camera length squared for a series of twist angles and $\alpha$ choices. The two smallest of the three twist tables are labeled, with the third implicit angle being the sum of these other two. \textbf{(3c)} To-scale illustrations of the overlap regions expected at a series of $\alpha$ choices for a hBN, bilayer graphene hetero-structure with inter-layer twists of 2 degrees, 15 degrees, and 17 degrees between the two graphene layers and each graphene layer and the hBN respectively. }
\end{figure*}

\section{Resolution-Driven Restrictions on Sample Morphology}

One major resolution limiting consideration in previous studies was the effect of beam-width biasing, as previous works reported stacking order and strain averaged over illumination regions of approximately 1 to 1.5 nm. These probe sizes are also larger than those seen in other electron microscopy techniques \cite{jiang2018electron, chen2020mixed, yang2024local, pelz2017low}, which often use probe widths on the order of 0.2-0.5 nm and also analyze the sample variability within illumination regions to obtain sub-angstrom resolution. Unlike many microscopy techniques, the minimum viable spatial resolution of BI is dependent on the sample morphology. To understand why we first note that the full-width at half maximum of the point spread function, or the modulo square of Fourier transform of $\exp(i\chi)$ where $\chi$ is the Scherzer aberration function \cite{scherzer1949theoretical}, and $\alpha$-dependent phase shift introduced by the lens, yields a typical measure for probe width. \cite{kirkland1998advanced} For many applications, the maximum tolerable phase error is then used to determine the maximal semi-convergence angle ($\alpha$) and its associated probe width of approximately 0.61 $\lambda/\alpha$ where $\lambda$ is the wavelength of the electron probe (Figure 1e). \cite{weyland2020tuning} We will not consider the effect of this phase error on the CBED pattern signal to noise as it is beyond the scope of this work, instead focusing on $\alpha < 10$ mrad where it is deemed tolerable with sufficient aberration correction \cite{weyland2020tuning} and consider the number of pixels we can obtain at each $\alpha$ the practical limitations associated with their extraction. \color{black} We note that in this discussion we consider the information limit associated with fitting the average intensity within overlapping Bragg disks, instead of the information limit set by the detector pixel size which may be realized in principle by considering the internal structure of these regions. However, we note that dark field intensities do not provide all of the phase information contained in the bright field. A larger resolution may also be obtained in theory through oversampling. \color{black}

In BI, we often cannot realize the optimal $\alpha$ as the large convergence angle will lead to an overlap of additional Bragg disks and prohibit the selective investigation of desired interfaces. To see how these considerations translate to the obtained spatial resolution, we first note that the probe wavelength we use is associated with a relatively low 80 kV accelerating voltage, which increases the probe width. The use of lower accelerating voltages is however more necessary in thin samples to reduce knock-on damage \cite{kretschmer2020formation, bell2014successful, egerton2012mechanisms}. The resulting probe width provides us with a measure of our spatial resolution, where we can assess changes in average displacement down to a minimum spatial resolution of roughly d if not slightly smaller \cite{james1999practical}. This can be used to determine the minimum $\alpha$ needed at each twist angle to resolve stacking order domain size changes to a given precision. For this, we define the AA stacking domains to include stacking configurations of $|\textbf{u}_{xy}| = 0 \pm 0.25 a_0$ so that a rigidly twisted structure has AA domains of width $0.25a_0/\sin(\theta/2)$. The $\alpha$ needed to resolve domain contractions to a given percent of the rigid value are shown in Figure 3a. We note that the BI methodology is most suitable for relatively small twist angles where dramatic changes to the AA domains are seen. 

For many sample morphologies, $\alpha$ must be smaller than 10 mrad. To see this, we show a series of anticipated overlap regions (Figure 3b), obtained using the relations illustrated in Figure 1a-b. We can then straightforwardly compute the number of overlap pixels (in units of the camera length squared) at a given choice of sample morphology and $\alpha$. We also note that small overlap regions are difficult to isolate in practice. This currently limits the spatial resolution and the range of samples that can be investigated with this technique. However larger $\alpha$ and therefore higher spatial resolution are possible than those used in previous works ($\alpha \approx 2-3$ mrad) \cite{kazmierczak2021strain, van2023rotational} if care is taken to define precise virtual apertures of variable shape (Figure 3c) instead of avoiding overlap with encapsulating layers entirely. We note that for many samples this will require virtual apertures to be defined separately for each pixel (or at least local regions of real space) instead of from diffraction patterns averaged over larger domains as was done previously \cite{kazmierczak2021strain, van2023rotational, craig2024local} and that precise automated determination of these small overlap regions is difficult for the same reason that the precise disk locations in these individual CBED patterns are hard to extract. This limitation may also be addressed by fitting regions associated with a greater number of overlapping disks to more complex expected expressions so that larger virtual apertures can be used, although such an approach requires improvements in the computational post-processing to be discussed. 


\section{Intensity Fitting To Determine Inter-layer Stacking}

We compare multiple fitting procedures to obtain the displacements $\textbf{u}_{xy}$ prescribed by equation 2. One approach we use involves fitting $\textbf{u}_{xy}$ and the coefficients $A_i, B_i$ sequentially, such that $\textbf{u}_{xy}$ at each pixel location $(x,y)$ and the coefficients for each of the twelve disks can each be fit independently and parallelized. In addition to being much faster and allowing for the coefficients to be optimized using linear least squares, we found that this use of alternating optimization \cite{bezdek2003convergence} was also more robust to experimental noise and permitted the use of many initial guesses at each iterative step, helping to avoid local minima. 

In this iterative approach, we first determine the optimal displacement vector $\textbf{u}_{xy}$ independently at each pixel (x,y) provided the fixed coefficients $A_i, B_i$. The experimental intensities are normalized, and $A_i=0, B_i=1$ are assumed. We used a dense uniform grid of 36 initial starting conditions to decrease the chance of obtaining local minima and constrain the values of $\textbf{u}_{xy}$ to reside within a single unit cell such that $\textbf{u}_{xy} = c^{(1)}_{xy} \textbf{a}_1 + c^{(2)}_{xy} \textbf{a}_2$ with $|c^{(1)}_{xy}| \leq 1/2$ and $|c^{(2)}_{xy}| \leq 1/2$ in terms of the separately measurable in-plane lattice vectors $\textbf{a}_1$ and $\textbf{a}_2$. This results in the following over-determined equation with an easily obtainable Jacobian, which we solved using Newton non-linear least squares independently for each pixel (x,y). \color{black} Here we use $(j,k)$ in place of the compound index $i$ to more clearly define the different Bragg disk locations in terms of the lattice vectors, such that $\textbf{g}_i = j \textbf{b}_1 + k \textbf{b}_2$ in terms of the standard 2D reciprocal lattice vectors $\textbf{b}_1, \textbf{b}_2$ associated with $\textbf{a}_1$ and $\textbf{a}_2$. \color{black}

\begin{eqnarray}
\begin{bmatrix} 
I_{10xy} \\
... \\
I_{jkxy} \\
\end{bmatrix}
&= \begin{bmatrix}
A_{10} + B_{10} cos^2(\pi c^{(1)}_{xy}) \\
... \\
A_{jk} + B_{jk} cos^2(\pi (jc^{(1)}_{xy} + kc^{(2)}_{xy})) \\
\end{bmatrix}
\end{eqnarray}

\color{black} Using the resultant $\textbf{u}_{xy} = c^{(1)}_{xy} \textbf{a}_1 + c^{(2)}_{xy} \textbf{a}_2$ values, \color{black} we determine the optimal coefficients ${A_i, B_i}$ independently for each of the twelve Bragg disk intensities indexed by i at each pixel $(x,y)$ using linear least squares. Fitting the coefficients at a fixed choice of $\textbf{u}_{xy}$ results in the following linear equation for each diffraction disk i. We note that the coefficients $A_i$, $B_i$ are assumed to be independent of pixel location over the N+1 by N+1 pixel field of view to avoid over-fitting, and can be assumed to be approximately 1 and 0 respectively following normalization of the intensities. 

\begin{eqnarray} 
\underbrace{
\begin{bmatrix}
    cos^2(\pi \textbf{g}_i \cdot \textbf{u}_{00} ) & 1 \\
    cos^2(\pi \textbf{g}_i \cdot \textbf{u}_{10} ) & 1 \\
    ... \\ 
    cos^2(\pi \textbf{g}_i \cdot \textbf{u}_{NN} ) & 1 \\
\end{bmatrix}}_{A_{coef}}
\cdot
\begin{bmatrix}
    A_i \\
    B_i \\
\end{bmatrix} 
= \underbrace{ \begin{bmatrix}
   I_{i00} \\
   I_{i10} \\
   ... \\
   I_{iNN} \\
\end{bmatrix} }_{b_{coef}}
\end{eqnarray}

The fitting of ${A_i, B_i}$ and $\textbf{u}_{xy}$ are then iterated until convergence. We compare this staged optimization to a joint optimization in which the coefficients and $\textbf{u}_{xy}$ are optimized simultaneously following one iteration of the same multi-start \textbf{u}$_{xy}$ optimization with fixed $A_i=1, B_i=0$. The mean |$\textbf{u}_{xy}$| obtained over the full set of pixels is shown for both approaches in Fig 4a for artificial data corresponding to a rigidly twisted structure with disk intensities that exactly obey equation 2 with independent Poisson noise associated with acquisition included for each disk intensity. Representative experimental datasets are shown in Fig 4c, showing that the extent of noise included is comparable or greater than that expected of weakly scattering samples like graphene and much greater than that expected of transition metal dichalcogenides (both acquired using $\alpha$ = 1.7 mrad). Fig 5a shows that the iterative approach consistently achieves a lower error in the displacements $\textbf{u}_{xy}$ (Fig 4a), \color{black} however this may not be robust when considering subtle differences in the initial conditions. Despite this, the increased speed and ability to use many initial guesses at each interactive step are still beneficial. \color{black} Avoidance of local extrema and noise mitigation are both active areas of research and alternative algorithms such as genetic algorithms \cite{mirjalili2019genetic} could prove better than this iterative multi-start approach, but were not required for experimental data of this nature.  

\begin{figure*}
\includegraphics[scale=0.85]{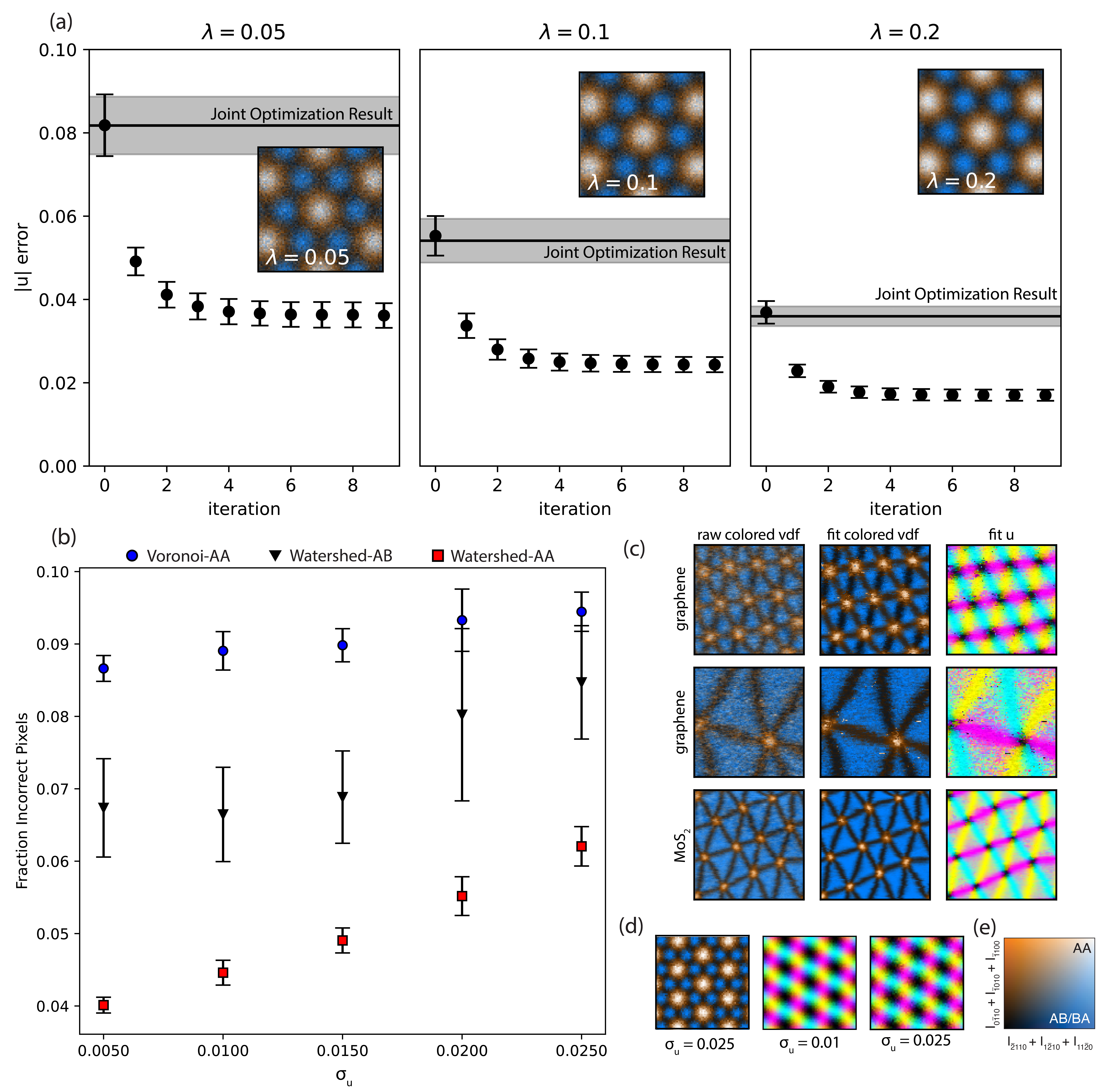}
\caption{ \textbf{(4a)} Mean $|\textbf{u}_{xy}|$ errors (in units of $a_0$) obtained through fitting intensities in the form of equation 2 (with A=1, B=0) using the iterative and joint optimizations described. Poisson noise is included by drawing each disk intensity independently from the Poisson distribution $P(k;\lambda I)=(\lambda I)^k\exp(-\lambda I)/k!$ where I is the noise-free intensity. A range of specified $\lambda$ choices are shown, which in practice will depend on the electron dose. Noise mitigation is expected when averaging over many pixels to obtain overlap regions, increasing the effective $\lambda$. The resulting signal-to-noise ratio of the interferometry pattern is shown using a colored virtual dark field in the inset. Error bars represent the standard deviation using 25 datasets randomly generated in this manner. The result of the joint optimization is shown as a solid line and its corresponding standard deviation as the greyed region. \textbf{(4b)} The fraction of incorrectly assigned pixels using each of the three described unwrapping procedures applied to the displacement fields expected of a rigidly twisted structure (shown in 4d) with Gaussian noise of zero mean and the specified standard deviation $\sigma_u$ (in units of $a_0$). Error bars represent the standard deviation using 25 datasets randomly generated in this manner. We note that pixels on the edge of the field of view can be incorrectly assigned even without noise in these approaches. \textbf{(4c)} Colored virtual dark fields (see Fig 4e) and displacement field maps (visualized using the color scheme defined in Fig 1c) illustrating the signal-to-noise ratios and fit efficacy of representative experimental data. \textbf{(4d)}  Colored virtual dark fields and displacement field maps (visualized using the color scheme defined in Fig 1c) for artificial data assuming a rigidly twisted structure and given Gaussian noise, as used in the comparison of the unwrapping approaches in (4b). \textbf{(4e)} Color legend for virtual dark fields illustrating qualitative atomic stacking classifications. \cite{craig2024local} } \end{figure*}

\section{Displacement Unwrapping for Strain Extraction}

The displacements obtained in the fitting procedure cannot yet be differentiated to assess local strain. This is because the obtained $\textbf{u}_{xy}$ are not uniquely determined by the data, since adding integer multiples of the lattice vectors $\textbf{a}_1$ and $\textbf{a}_2$ to $\textbf{u}_{xy}$ results in the same set of intensities due to the sample's symmetry. For centro-symmetric materials, flipping the sign of $\textbf{u}_{xy}$ also results in the same pattern. Owing to these ambiguities, the raw displacement vectors obtained are not smoothly oriented and cannot be naively differentiated, even without experimental noise or imperfections in the fitting procedure. To resolve this, we can determine the sign and lattice vector offsets needed to obtain smoothly varying displacements. While not specific to BI-obtained displacements, such a procedure is not needed when analyzing simulated data with uniform moir\'{e} domains as the effect of lattice symmetry can be accounted for by easily removing low-frequency modes. Additionally, this degeneracy does not feature in more conventional strain-mapping procedures that monitor gradual shifts in the Bragg disk locations. The problem will however generally apply when attempting to differentiate experimentally obtained displacements, given that the extent of noise is large enough that we cannot naively choose locally optimal offsets.

\color{black} We note that this physical degeneracy is from a phase ambiguity in the projected electrostatic potential, and that phase ambiguities such as these feature prominently across a wide array of physical techniques, from remote sensing to magnetic resonance imaging. A great deal of previous research has attempted to resolve these degeneracies through phase unwrapping procedures, often but not always cast as optimizing a non-local cost function.\cite{fienup1982phase, shechtman2015phase, goldstein1988satellite, sun2018geometric, jaganathan2016phase, judge1994review,schofield2003fast} These existing strategies for phase unwrapping we are aware of primarily find a smoothly varying scalar quantity $\phi^{unwrap}_{xy} = \phi_{xy} + 2\pi n_{xy}$ given $\phi_{xy}$ with $n_{xy} \in Z$, inverting $\tan^{-1}(\sin(x)/\cos(x))$ to resolve the argument degeneracy of $\cos(x)$ or similar. 

We seek instead to resolve a degeneracy in $\cos^2(\textbf{x} \cdot \hat{g}_i)$, therefore unwrapping a vector quantity along directions set by normalized reciprocal lattice vectors $\hat{g}_i$. Importantly there is also an additional sign degeneracy in a squared cosine, so the strategies developed for interpreting phases proportional to $\textbf{u} \cdot \hat{g}_i$ in geometric phase analysis \cite{hytch1998quantitative} or general 2D phase unwrapping techniques \cite{ghiglia1996minimum, ghiglia1994robust, spagnolini19952} cannot be used directly. This sign degeneracy is a manifestation of a more general issue involving a loss of phase information in overlapping dark field reflections compared to the bright-field Fourier spectrum. 

So since the degeneracy here is a more complex manifestation of the lattice symmetry, we develop an application-specific approach for which we can use physical intuition regarding sample morphology to reduce computational cost. \cite{van2023rotational, kazmierczak2021strain} Future work adapting more general phase unwrapping approaches to this application may be promising. We also note that since we only care for the derivative of the displacement, it may be possible in principle to avoid unwrapping altogether by modifying the finite difference stencil as done in essentially non-oscillatory schemes. \cite{cockburn1998essentially, liu1994weighted} We found this strategy sensitive in practice to noise. \color{black}

Thus to unwrap the displacement field and obtain moiré-scale strain, we seek to find the optimal $s_{xy}, n_{xy}, m_{xy}$ at each pixel location $(x,y)$ given $\textbf{u}_{xy}$ such that the (squared) euclidean distance of $\textbf{u}^{\rm{unwrap}}_{xy}$ from all four of its neighbors is minimized, thus minimizing a quadratic cost function to variations in integer parameters. \cite{van2023rotational}

\begin{align}
\textbf{u}^{\rm{unwrap}}_{xy} = s_{xy} \textbf{u}_{xy} + n_{xy} \textbf{a}_1 + m_{xy} \textbf{a}_2 \\ n_{xy},m_{xy} \in Z, s_{xy} = \pm 1
\end{align}

Liberal use of integer quadratic programming is prohibitively expensive, so we considerably pre-process the data so that it need only be applied in small regions. We first pre-processed the data using physical intuition regarding its geometry, partitioning the displacements into zones of constant lattice vector offsets $(n_r,m_r)$, such that $n_{xy}, m_{xy} = n_r, m_r$ for all pixels $(x,y)$ in region r. \cite{van2023rotational}

\begin{figure}
\includegraphics[scale=0.8]{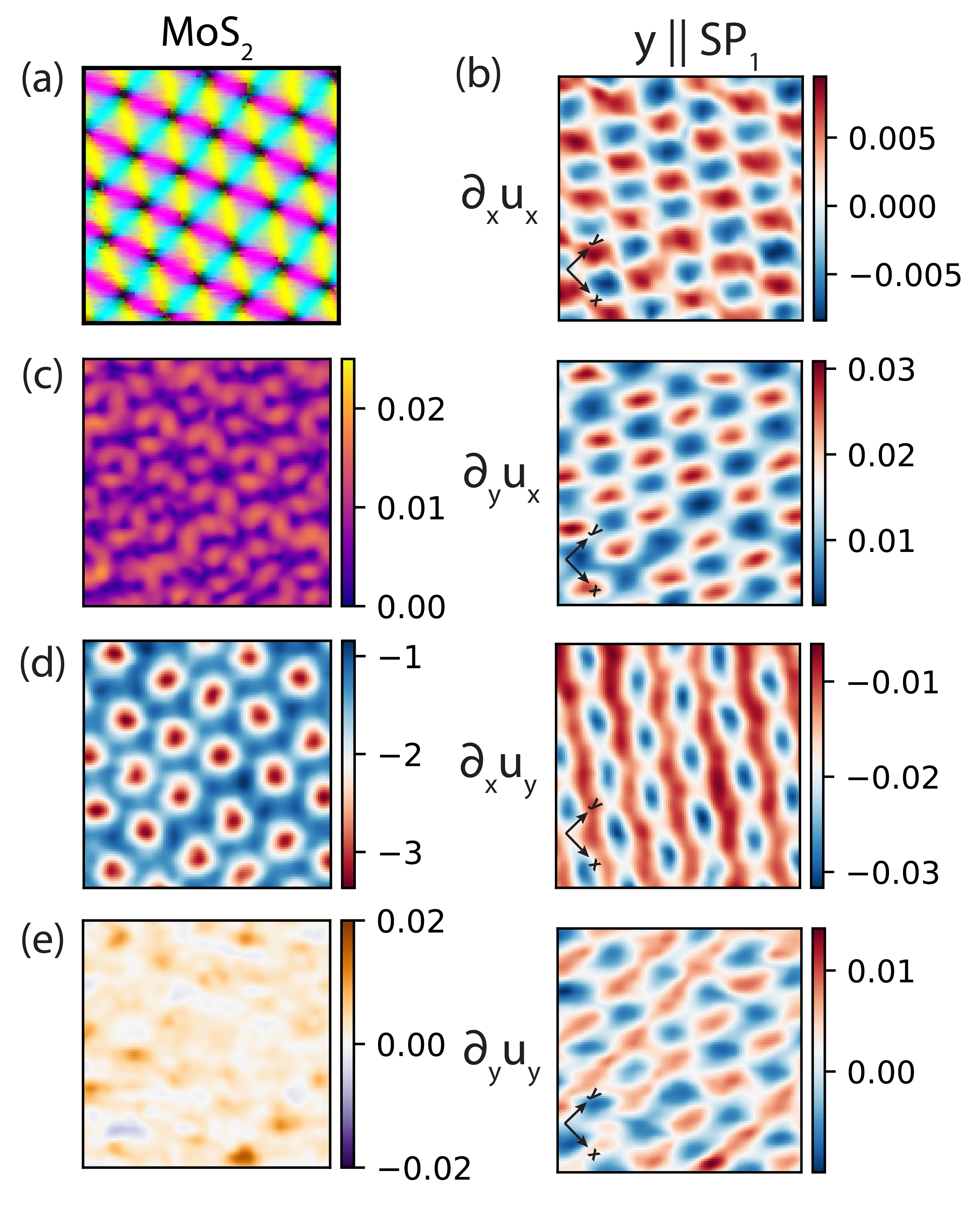}
\caption{ \textbf{(5a)} Displacement field map (visualized using the color scheme defined in Fig 1c) for a representative experimental dataset of hBN encapsulated bilayer MoS$_2$ with a 1.77 degree twist, fit using the described iterative procedure. The field of view is 100 by 100 pixels, corresponding to 50 by 50 nm. \textbf{(5b)} Components of the strain tensor $\nabla \textbf{u}_{xy}^{\rm{top}}$ obtained by numerically differentiating the displacement field following unwrapping via the Watershed-AA approach. Values provided are unit-less after accounting for the difference in scale for the displacements ($a_0$ = 0.315 nm for MoS$_2$) and pixels (step size of 0.5 nm). Results are shown using a coordinate system where the y-axis is taken to lie along $SP_1$, necessary to obtain $\nabla \cdot \textbf{u}_{xy}^{\rm{top}} \approx 0$ and $\nabla \times \textbf{u}_{xy}^{\rm{top}} \approx \theta_t$ in rigidly twisted structures. \textbf{(5c-e)} The corresponding engineering sheer strain, local rotations $\nabla \times \textbf{u}_{xy}^{\rm{top}}$ (shown in degrees), and local dilation $\nabla \cdot \textbf{u}_{xy}^{\rm{top}}$}
\end{figure}

One partitioning approach we used involves the watershed algorithm. \cite{kornilov2018overview} In this method, several regions are specified (in this case the centers of the $|\textbf{u}_{xy}|\approx0$ (AA-type) stacking regions) and pixels of increasing $|\textbf{u}_{xy}|$ are successively assigned. We denote this as watershed-AA. The watershed-AA approach proved effective on the data sets with large, dispersed regions of $|\textbf{u}_{xy}|\approx0$ stacking separated by thin boundaries, as was seen for anti-parallel aligned twisted transition metal dichalcogenides \cite{van2023rotational}. We also investigate an alternative approach based on a Voronoi partition \cite{okabe2009spatial}, where region boundaries are drawn halfway between the AA locations. This method was successful for highly noisy data but failed when the moir\'{e} wavelength varied even marginally over the field of view. \cite{van2023rotational} Lastly, we investigate an approach where the watershed algorithm is applied to the angle of $\textbf{u}_{xy}$ instead of its magnitude to select Bernal-type stacking regions for structures that minimize $|\textbf{u}_{xy}|\approx0$ like large-angle twisted graphene and parallel aligned twisted transition metal dichalcogenides. We denote this \color{black} new approach as watershed-AB, which was not used in our prior applications, \cite{van2023rotational} but is conceptually similar. \cite{kazmierczak2021strain}  \color{black} After use of any of these three segmentation approaches, the values $n_r, m_r$ are determined based on the connectivity of neighboring regions and $s_{xy}$ is chosen to maximize a local curl. A quadratic integer program can then be applied to the pre-processed data. \cite{van2023rotational} 

We compare these three segmentation approaches on artificial $\textbf{u}_{xy}^{\rm{unwrap}}$ data corresponding to a rigidly twisted structure with Gaussian-distributed noise to model the effects of sample defects and contamination, which is then converted into $\textbf{u}_{xy}$ following Eqn 4. Fig 4b shows the fraction of incorrectly assigned pixels after each segmentation approach. These results illustrate that the Watershed-AA approach performs best for the rigidly-twisted domains, but its performance declines more rapidly with increasing noise than the Voronoi approach. The Watershed-AB approach is more variable to subtle details in the applied noise, which can impact the angle of $\textbf{u}_{xy}$ more easily than its magnitude. For applications to experimental data, the most appropriate choice will depend heavily on the morphology of the moir\'{e} pattern.


Following the unwrapping procedure, $\textbf{u}_{xy}^{\rm{unwrap}}$ can then be differentiated numerically, for which we use a centered finite difference stencil. We note that the measured displacements are cell-averaged quantities, which some other stencil choices would need to account for and a factor of one half is introduced as we consider the intralayer strain $\nabla \textbf{u}_{xy}^{\rm{top}}$. In practice there is often a rotation between the real-space coordinate system and that used to define $\textbf{u}_{xy}$, which need be accounted for. \cite{kazmierczak2021strain, van2023rotational} 

Using the techniques described here, we can directly measure the moir\'{e}-scale strain in bilayer systems. Experimental strain distributions are shown in Fig 5 for MoS$_2$ with a twist angle of 1.77 degrees using a series of coordinate choices. The sheer strain, curl, and divergence maps are shown in Fig 5c-e, expected to be uniformly 0, $\theta_t$, and 0 in a rigid structure respectively. The clear local structure in 5c and 5d are consequences of atomic reconstruction, showing that MoS$_2$ distorts to increase the local twist angle within the $|\textbf{u}_{xy}|\approx 0$ (shown black in Fig 5a), reducing the size of these high-energy stacking regions. Simultaneously, the distortion decreases the local twist angle within the low-energy stacking regions (Fig 5d). These two rotational effects act in a manner that accumulates shear strain along their boundaries (Fig 5c) while the dilational strain associated with uniform stretching of the lattice remains negligible (Fig 5e). \cite{van2023rotational}

\section{Conclusion}

We have outlined and analyzed various numerical and instrumental limitations for the BI technique and presented solutions to generalize its applicability to a wider range of bilayer systems. Future work will aim to expand this method for use in the limit of smaller converged probe sizes and more general samples. This may involve any of the discussed strategies including the use of more precisely defined non-spherical virtual apertures to enable smaller CBED probes, fitting disk locations in tandem with their intensities, more involved post-processing techniques along the lines of those discussed for interpreting multi-layer overlap regions, and improvements to the algorithms used. 

\appendix


\begin{acknowledgments}
\color{black} We acknowledge helpful conversations with C. Groshner and formative contributions from N. Kazimerzak, K. Bustillo, H. Brown, J. Ciston, T. Taniguchi, and K. Watanabe in the referenced initial realization of this approach. \color{black} This material is based upon work supported by the US National Science Foundation Early Career Development Program (CAREER), under award no. 2238196 (D.K.B). I.M.C. acknowledges a pre-doctoral fellowship award under contract FA9550-21-F-0003 through the National Defense Science and Engineering Graduate (NDSEG) Fellowship Program, sponsored by the Air Force Research Laboratory (AFRL), the Office of Naval Research (ONR) and the Army Research Office (ARO). Work at the Molecular Foundry was supported by the Office of Science, Office of Basic Energy Sciences, of the U.S. Department of Energy under Contract No. DE-AC02-05CH11231. 
\end{acknowledgments}

\section*{Conflict of Interest}
The authors have no conflicts to disclose.

\section*{Data Availability Statement}
The data that support the findings of this study are available within the article. \color{black} An python implementation of the discussed approach is available at github.com/bediakolab/pyInterferometry, which imports some functionality from py4DSTEM \cite{savitzky2021py4dstem} among other open-source modules such as GEKKO. \cite{beal2018gekko} \color{black}

\providecommand{\noopsort}[1]{}\providecommand{\singleletter}[1]{#1}%

\end{document}